\documentstyle[12pt,aaspp4]{article}

\def\simlt{\ {\raise-.5ex\hbox{$\buildrel<\over\sim$}}\ }
\def\simgt{\ {\raise-.5ex\hbox{$\buildrel>\over\sim$}}\ }

\def\asca{{\it ASCA\/}}
\def\chandra{{\it Chandra\/}}
\def\rosat{{\it ROSAT\/}}
\def\pss{{\sc pss\/}}
\def\pimms{{\sc pimms\/}}
\def\xspec{{\sc xspec\/}}
\def\ginga{{\it Ginga\/}}

\begin{document} 


\title{New X-ray Constraints on Starburst and Seyfert
Activity in the Barred Spiral Galaxy NGC~1672}

\author{P. J. de~Naray,$^1$ W. N. Brandt} 
\affil{Department of Astronomy \& Astrophysics, The Pennsylvania State
University, 525 Davey Lab, University Park, PA 16802}


\author{J. P. Halpern}
\affil{Columbia Astrophysics Laboratory, Columbia University, 538
West 120th Street, New York, NY 10027}

\author{K. Iwasawa}
\affil{Institute of Astronomy, Madingley Road, Cambridge CB3 0HA
U.K.}


\footnotetext[1]{NASA-supported undergraduate research associate.}

\setcounter{footnote}{1}


\begin{abstract}
The nearby barred spiral galaxy NGC~1672 shows dramatic starburst 
activity and may also host a Seyfert~2 nucleus. We present new 
X-ray observations that set constraints on starburst and Seyfert activity 
in NGC~1672.
Two \rosat\ HRI exposures, taken in 1992 and 1997, are used to 
investigate long-term variability of the known X-ray sources and 
to search for new sources of X-ray emission. We find 
large-amplitude ($\approx 69$\%) variability from X-3, one of 
the off-nuclear sources located near an end of the galactic bar. 
X-3 has a peak observed 0.2--2.0~keV luminosity of 
$\approx 2.5\times 10^{39}$~erg~s$^{-1}$, and it is probably a luminous 
X-ray binary or young supernova remnant. We do not observe variability 
of the nuclear source X-1 or the strong off-nuclear source X-2.
Our analyses also reveal two new off-nuclear sources, one of which 
is associated with a bright region along a spiral arm, and we 
find evidence for large-scale diffuse X-ray emission throughout 
part of the disk of NGC~1672. 

Furthermore, we use \asca\ data taken in 1995 to constrain the 
hard X-ray properties of NGC~1672. While the nuclear source X-1 
is the dominant soft X-ray source in NGC~1672, we find that the
bulk of the 2--10~keV and 5--10~keV emission is spatially 
coincident with the off-nuclear source X-3, giving it an
apparent 0.2--8~keV luminosity of $\simgt 6\times 10^{39}$~erg~s$^{-1}$. 
A power-law plus Raymond-Smith model provides an acceptable
fit to the full-band \asca\ spectra. We do not find any evidence for 
a luminous but absorbed nuclear X-ray source. If there is a luminous 
Seyfert~2 nucleus in NGC~1672, it must be obscured by a `Compton-thick' 
torus with a column density of $\simgt 2\times 10^{24}$~cm$^{-2}$. 
\end{abstract}



\section{Introduction}

NGC~1672 is a $V=10.1$ barred spiral galaxy in the Southern 
hemisphere that is viewed nearly face on. It shows dramatic nuclear 
and extranuclear star-formation activity including starburst regions 
near the ends of its strong bar (e.g., Pastoriza 1967; 
Osmer, Smith \& Weedman 1974; Baumgart \& Peterson 1986). It may 
have a Seyfert~2 nucleus (e.g., V\'eron, V\'eron \& Zuiderwijk 1981; 
Brandt, Halpern \& Iwasawa 1996, hereafter BHI96), but at present 
this has not been firmly established. NGC~1672 is at a distance of 
16.3~Mpc ($H_{0}=70$~km~s$^{-1}$~Mpc$^{-1}$), and the Galactic 
neutral hydrogen column density in its direction is 
$N_{H}=(2\pm 1)\times 10^{20}$~cm$^{-2}$ (Heiles \& Cleary 1979). 

X-ray emission is powerful for probing both the endpoints of stellar 
evolution and Seyfert activity, and NGC~1672 is known to be a 
prodigious source of X-rays. BHI96 presented a detailed \rosat\  
Position Sensitive Proportional Counter (PSPC) and High Resolution 
Imager (HRI) study of NGC~1672, and they found several sources of X-ray 
emission coincident with the galaxy. 
The nucleus showed the strongest emission. This emission was 
slightly resolved, suggesting at least some starburst X-ray 
activity, but the \rosat\ data were unable to determine whether the bulk 
of the emission was from a starburst region or hidden Seyfert. 
Awaki \& Koyama (1993) claimed a hard X-ray detection of a 
$L_{2-10~{\rm keV}}\approx 10^{41}$~erg~s$^{-1}$ Seyfert nucleus 
with the \ginga\ Large Area Counter (LAC), but concerns 
about source confusion were present (the LAC was a non-imaging detector 
and had a $2^\circ\times 4^\circ$ field of view).
In addition to the nuclear X-ray emission, BHI96 also found luminous 
X-ray sources coincident with the starburst regions near the ends of 
the galactic bar. It was difficult to determine the nature of these 
sources due to limited photon statistics. They could be consistently 
modeled as superbubbles, 
collections of X-ray binaries, or
`super-Eddington sources' (e.g., Fabbiano 1998).  

To investigate the X-ray properties of NGC~1672 further, we performed 
a second \rosat\ HRI observation of it in 1997. Our  
goals for this observation were (1) to study the long-term X-ray variability
properties of the currently known sources, (2) to search for new X-ray 
sources that were not detected before due to their variability, and 
(3) to obtain better photon statistics for investigation of any
weak point sources or large-scale diffuse X-ray emission. 
Here we present the results from our second HRI observation as well
as an improved analysis of the original 1992 HRI observation. 
We also use \asca\ data from 1995 to constrain the hard 
X-ray properties of NGC~1672. 


\section{\rosat\ HRI Observations, Data Reduction, and Analysis}

\subsection{Observations and Basic Analysis}

NGC~1672 was observed with the \rosat\ HRI (e.g., David et~al. 1999)
during the period 27~January~1997 to 2~March~1997 (35.0~ks exposure time).  
The observation went smoothly, and reduction and analysis of the 
data were performed with the {\sc asterix} software system 
(Allan \& Vallance 1995). 
We created three images for subsequent analysis: 
(1) an image from the 1997 HRI observation,
(2) an image from the original 1992 HRI observation (24.5~ks; see BHI96), and
(3) a `combined' image using the data from both the 1992 and 1997 observations 
(59.5~ks of total exposure time). 
All images were made with $2^{\prime\prime}$ per pixel resolution so 
that the HRI point spread function (PSF) was oversampled by a factor
of $\approx 2$. 
We also used only HRI channels 3--8, since this channel range provides
maximum sensitivity for detecting point sources and 
low-surface-brightness diffuse emission (see \S3.4 of David et~al. 1999). 

\subsection{Point Source Searching, Parameterization, and Variability}

The three images described in \S2.1 were analyzed using
the point source searching software {\sc pss} (Allan 1995). 
\pss\ locates point sources by convolving an HRI image with a 
position-dependent model for the HRI PSF, and it calculates
point source significances using the Cash (1979) statistic. 
We used constant backgrounds for each of the three images, and
our manual background analyses showed that this is valid
in the region around NGC~1672. We consider for further study 
point sources that are within $2.3^\prime$ of the center of
NGC~1672 (this corresponds to the main optical extent of the 
galaxy) and are found to be significant at $\geq 4\sigma$ 
in at least one of the three images. Given these criteria, we
expect $\approx 0.3$ false sources in total. 
\pss\ also creates a `significance map' showing the HRI image 
after it has been convolved with the PSF model; we show
the significance map for the 1997 observation in Figure~1.

In Table~1 we give (background-subtracted) source counts, mean count 
rates, fluxes, and luminosities for all sources. We have adopted 
the source naming convention of BHI96 but have also added sources 
X-7, X-8 and X-9.\footnote{Sources X-4 and X-6 of BHI96 are omitted 
from Table~1 and all subsequent discussion because they are outside 
the main optical extent of the galaxy and the regions of strongest 
star formation. In addition, X-6 is probably associated with a bright 
foreground star.}
Source counts were calculated using circular regions chosen to fit the 
morphology of each source. When sources were not detected in one of 
the images, we calculated $3\sigma$ upper limits. 
We used the \pimms\ (Mukai 1997) software package to convert 
the mean count rates for each source into 0.2--2.0~keV fluxes
and luminosities. For X-1, we adopted the Raymond-Smith thermal 
plasma model of BHI96 to perform this conversion. For all other
sources, we adopted a power-law model with a photon index of 
$\Gamma=2$ and the Galactic column density. 

The 1997 observation has revealed a new probable X-ray source (X-8) 
within the optical extent of NGC~1672. This source lies near the edge
of one of the prominent spiral arms. It is detected at the  $4.0\sigma$ 
level, so there is a small but non-negligible chance that it is due
to a statistical fluctuation. 
In addition, our analyses of all three images show that the extended 
source X-3 of BHI96 actually appears to be two separate sources
(labeled here as X-3 and X-7; see Figures~1 and 2). This 
difference between BHI96 and our analysis is understandable
because our HRI data have been through a revised 
processing that includes better filtering as well as satellite
aspect information from the \rosat\ Wide Field Camera 
(M. Corcoran 1998, private communication). These factors improve
the spatial resolution. In addition, BHI96 used the full
HRI band while we only use channels 3--8 (see \S2.1). 
Finally, searching the 59.5~ks combined image reveals another
new point source, X-9 (see Figure~2). This X-ray source is spatially 
coincident with a bright region along one of the spiral 
arms that contains three large H~{\sc ii} regions
(numbers 31, 34 and 39 in Table~2 of Evans et~al. 1996). The
alignment of the X-ray and optical emission supports the reality
of the X-ray detection. 

To check the X-ray spatial extent results of BHI96, we have performed
spatial extent analyses for X-1 and X-2 using both the 1992 and 1997 
observations. Our results are in general agreement with those
of BHI96: X-2 does not show any clear evidence for spatial extent
while X-1 is extended on about the scale of the nuclear ring. 
X-1 is more extended than X-2, and this strongly suggests that
its apparent extent is not an artefact of aspect solution errors
(see Morse 1994). 
 
We have compared the mean count rates between the 1992 and 1997 observations 
to search for long-term source variability.\footnote{The effective area of 
the HRI has been stable to within $\approx 5$\% during this time period 
(see \S2.3.3 of David et~al. 1999).} X-3 shows significant mean count rate 
variability with an $\approx 69$\% increase between 1992 and 1997. The 
Poisson probability that this variability is merely due to
a statistical fluctuation is $<0.5$\%. We have searched for rapid 
variability of X-3 within the 1997 data, and we do not find any 
highly significant variability. 
The source X-5, which is thought to be associated with a foreground star 
(see BHI96), also shows significant variability.\footnote{On 
18 October 1998 we obtained optical spectra of this star with the 
1.5-m Ritchey-Chretien telescope at the Cerro Tololo Inter-American 
Observatory. It appears to be a G5 main-sequence star with $V=12.5$, 
and it is thus at a distance of $\approx 300$~pc. The observed 
X-ray-to-optical flux ratio is consistent with that expected for a G5 
star (see Figure~1 of Maccacaro et~al. 1988), and the observed X-ray 
variability is not implausibly large for such a star.}
X-1 and X-2 do not show significant variability. We place upper limits 
of $\approx 15$\% and $\approx 30$\% on the variability of X-1 and X-2,
respectively. 


\subsection{Diffuse X-ray Emission}

We have used the adaptive kernel smoothing algorithm {\sc asmooth} 
(Ebeling, White \& Rangarajan 1999) to search for diffuse X-ray 
emission in the 59.5~ks combined image. {\sc asmooth} convolves 
a Gaussian of variable, position-dependent size with an image to 
allow detection of spatial structure on a wide range of scales, and 
it provides a quantitative assessment of the significances of 
observed features.
We find evidence for an $\approx 0.5^\prime\times 1^\prime$ area 
of diffuse X-ray emission between 
X-1 and X-2 (see Figure~3). We have examined if this emission 
can be explained as `spill over' from X-1 and X-2, and the observed
emission appears to significantly exceed that expected from these two
sources (even after taking into account reasonable \rosat\ aspect solution
errors; see Morse 1994). Note that similar emission is not seen between
X-1 and X-3 even though these two sources are closer together than X-1 and 
X-2 ($1.1^\prime$ versus $1.4^\prime$ separation). {\sc asmooth} indicates that 
the diffuse emission is significant at the $\approx 3.7\sigma$ level, and our 
manual calculations using rectangular regions give similar significances. 
We estimate $\approx 40$ counts in total from the diffuse emission, so 
the count rate is $\approx 7\times 10^{-4}$~count~s$^{-1}$. This 
emission may be associated with hot, diffuse gas in the interstellar
medium of NGC~1672. If we adopt a 0.55~keV Raymond-Smith model 
with solar abundances and the 
Galactic column density, we estimate a total 0.2--2.0~keV flux of 
$\sim 2\times 10^{-14}$~erg~cm$^{-2}$~s$^{-1}$ and a 0.2--2.0~keV 
luminosity of $\sim 6\times 10^{38}$~erg~s$^{-1}$
(the flux and luminosity values depend fairly strongly upon 
the unknown abundances of the gas). 
Of course, we are not able to prove that this emission must be from
truly diffuse hot gas. It could also plausibly arise from several weak 
point sources that are individually below our detection threshold. 


\section{\asca\ Observations, Data Reduction, and Analysis}

\subsection{Observations and Basic Analysis}

NGC~1672 was observed with \asca\ (Tanaka, Inoue \& Holt 1994) on 
23--24 November~1995. Both Solid-state Imaging Spectrometer CCD
detectors (SIS0 and SIS1) and both Gas Imaging Spectrometer 
scintillation proportional counters (GIS2 and GIS3) were operated. 
The SIS were operated in 1~CCD mode, and the GIS were operated in 
PH mode. We have used the Revision~2 processed data from 
Goddard Space Flight Center for the analysis below, and we have 
adopted the standard Revision~2 screening criteria (see Pier 1997).
After data screening, the exposure times were 33~ks for SIS0/SIS1 
and 40~ks for GIS2/GIS3. Data reduction was performed using 
{\sc xselect} (Ingham \& Arnaud 1998). 

The spatial resolution of \asca\ is a significant limiting factor
when studying NGC~1672, so we will present only the main results of
our \asca\ imaging analysis. The PSFs of the \asca\ X-ray Telescopes 
have half-power diameters of $\approx 3^{\prime}$ (compare with
Figures~1--3 and note that X-1 and X-3 are only separated by
$1.1^\prime$). However, they also have fairly sharp cores that 
allow two point sources as close as $\approx 30^{\prime\prime}$ 
to be resolved. In the case of the GIS, the final spatial resolution 
is worsened somewhat further by the detectors' spatial responses,
and the GIS cannot effectively study structure on scales smaller 
than $\approx 1^\prime$. 
%
%
For sources observed in an approximately on-axis manner, the \asca\ 
SIS and GIS positional uncertainties are $\approx 40^{\prime\prime}$
and $\approx 50^{\prime\prime}$, respectively (90\% confidence; 
Gotthelf 1996 and E. Gotthelf 1999, private communication). 

We have created 
full (0.6--10~keV for SIS and 1.0--10~keV for GIS), 
soft (0.6--2.0~keV for SIS and 1.0--3.0~keV for GIS), 
hard (2--10~keV for SIS and GIS), and
ultra-hard (5--10~keV for SIS and GIS) images for each of the four
\asca\ detectors. We also created summed SIS0+SIS1 and GIS2+GIS3 images 
in all bands. These images will be used in the spatial analysis below.
SIS images were made with $6.3^{\prime\prime}$ per pixel resolution, 
and GIS images were made with $15.0^{\prime\prime}$ per pixel resolution. 

\subsection{Spatial Analysis}
 
We have used {\sc ximage} (Giommi, Angelini \& White 1997) and {\sc idl}
to inspect the images created in \S3.1. At low energies our results are
consistent with those expected based upon the \rosat\ data. 
X-1 is observed to be the dominant X-ray source in NGC~1672 (see Figure~4a),
and we also detect X-3 and probably X-2 in the SIS images. We have
compared the \asca\ and \rosat\ data to examine source variability. 
We do not find any strong evidence for variability, although source 
confusion (see \S3.1) significantly limits our variability analyses. 

In the harder 2--10~keV and 5--10~keV images, however, NGC~1672 has a 
considerably different appearance (see Figure~4b). 
The dominant 2--10~keV emission is observed to be coincident with the HRI 
sources X-3 and X-7 (in both the SIS0+SIS1 and GIS2+GIS3 images), and we 
detect only weak 2--10~keV emission from X-1. Note that X-3 was also found 
to be the hardest source in the \rosat\ PSPC band by BHI96. 
In the 5--10~keV GIS2+GIS3 image, the only detectable emission from NGC~1672 
is coincident with X-3 and X-7. NGC~1672 is not detected in the 5--10~keV 
band by the SIS detectors, probably due to their smaller effective 
areas in this energy range (see Tanaka, Inoue \& Holt 1994). 


\subsection{Spectral Analysis}

Due to the limited spatial resolution of \asca\ (see \S3.1), we have 
only been able to perform reliable spectral analyses for NGC~1672 as 
a whole (see \S7.4 of {\it The ASCA Data Reduction Guide\/}). At 
energies above $\approx$~3--5~keV, X-3 is the dominant source, but
at lower energies our spectra contain contributions from X-1, X-2
and X-3. We have extracted spectra from each detector using all of the
acceptable exposure time, and we have only used data 
where NGC~1672 is reliably detected (0.6--5~keV for the SIS 
and 1--8~keV for the GIS). We have grouped 
our spectra so that there are 15 photons per spectral data point,
to allow the use of $\chi^2$ fitting techniques. We have first
performed preliminary spectral fitting for each of the \asca\ detectors
separately, and although the errors were large we obtained consistent
results. We have therefore jointly fitted the spectra
from all four detectors, and we detail these results below. Aside
from the absolute model normalizations, we tied together the fit 
parameters across the four \asca\ detectors. We use \xspec\ 
(Arnaud 1996) for all spectral fitting below. 

We start by fitting the data in the 2--8~keV band, where X-3
is the dominant source. We use a basic power-law model with the 
Galactic absorption column. This model provides an acceptable fit with 
$\chi^2=50.4$ for 56 degrees of freedom. We obtain a power-law photon 
index of $\Gamma=1.62^{+0.29}_{-0.40}$ (fit parameter errors are for 
$\Delta\chi^2=2.71$), and there is no evidence for systematic positive 
residuals near the 6.4--6.97~keV iron~K$\alpha$ lines. The 2--8~keV 
flux is $4.5\times 10^{-13}$~erg~cm$^{-2}$~s$^{-1}$, corresponding
to a 2--8~keV luminosity for NGC~1672 of
$1.4\times 10^{40}$~erg~s$^{-1}$. 
We have then applied this same model to the full 0.6--8~keV band.
We obtain a steeper power law with 
$\Gamma=2.06^{+0.13}_{-0.13}$ and $\chi^2=167.4$ for 153 
degrees of freedom. The fit is statistically acceptable, although
we observe systematic positive residuals from 0.6--1.0~keV and
systematic negative residuals from 1.1--1.6~keV. These are
understandable in the context of the \rosat\ spectral fitting
presented in \S2.3 of BHI96; the emission from NGC~1672 below
$\approx 2$~keV appears to be largely from hot, diffuse gas
that is associated with X-1, X-2 and X-3 (i.e., it is 
{\it not\/} primarily the diffuse emission discussed in \S2.3). 
If we take the best fitting 2--8~keV model and extrapolate it
downward in energy, we see clear positive residuals that are
likely to be associated with the thermal gas emission 
(see Figure~5). We note that such emission is commonly seen from
galaxies with starburst activity (e.g., Ptak et~al. 1999). 
We have added a Raymond-Smith component to our basic power-law model,
and this provides a highly significant improvement in overall fit
quality. We obtain a plasma temperature of $kT=0.66^{+0.16}_{-0.36}$~keV,
a photon index of $\Gamma=1.65^{+0.23}_{-0.25}$, and $\chi^2=139.0$ 
for 151 degrees of freedom ($\Delta\chi^2=28.4$ compared to the
basic power-law model; fit parameter errors are for $\Delta\chi^2=4.61$).
The plasma temperature from our fit is consistent with that of
BHI96, and the underlying photon index agrees better with that
seen in the 2--8~keV band. The derived plasma temperature is also
in good agreement with the temperatures obtained for other starburst 
galaxies (e.g., Figure~3 of Ptak et~al. 1999). The plasma abundances 
are poorly constrained, and we have adopted solar abundances in
the fit above. The derived plasma temperature, power-law photon
index, and X-ray luminosity do not depend sensitively upon the 
abundance assumptions for abundance enhancements up to a factor
of $\approx 5$, and we note that Storchi-Bergmann et~al. (1996) argue 
for moderately supersolar abundances in the nucleus of NGC~1672. 
In reality, we believe that the thermal emission below $\approx 2$~keV 
is likely to arise from gas with a range of temperatures and abundances,
and our single Raymond-Smith component model should not be taken too 
literally (see Ptak et~al. 1999 for further discussion on this matter). 
Comparison of the soft and hard X-ray luminosities of X-3 suggests
that this source may suffer from significant intrinsic absorption
(see \S4 for further discussion), and we have examined if our data 
constrain an absorption column affecting only 
the power-law spectral component. Unfortunately, our data 
do not tightly constrain such absorption, and column densities
ranging from 0~cm$^{-2}$ to $\approx 2\times 10^{22}$~cm$^{-2}$ are
allowed by our data.  
Due to significant \asca-\rosat\ cross-calibration uncertainties 
(e.g., Iwasawa, Fabian \& Nandra 1999) as well as the time variability 
of X-3, we have not performed detailed joint spectral fitting of the 
data from these two missions. 

The 2--8~keV flux we measure from NGC~1672 is $\approx 7$ times smaller 
than the 2--10~keV flux of $3\times 10^{-12}$~erg~cm$^{-2}$~s$^{-1}$
given by Awaki \& Koyama (1993), and our 
image analyses in \S3.2 show that most of the 
2--10~keV flux originates outside the nucleus. Thus, the \asca\ data 
do not provide evidence for a highly absorbed nuclear X-ray source 
as might be expected for a Seyfert~2. However, the \asca\ data do
not rule out a Seyfert~2 nucleus where the torus has a large 
enough column density to absorb almost all X-rays below 8~keV. 
Given the emission at other wavelengths (see \S3.1 of BHI96), 
a nuclear X-ray source with a luminosity up to 
$\approx 10^{43}$~erg~s$^{-1}$ appears plausible. 
We can estimate the required torus column density for a plausible scenario
using our spectral fits above. We have added an extra $\Gamma=2$ power 
law to our best 0.6--8~keV fit and assumed an unabsorbed 2--8~keV flux 
of $3\times 10^{-12}$~erg~cm$^{-2}$~s$^{-1}$ for this power law
(corresponding to a 2--8~keV luminosity of $\approx 10^{41}$~erg~s$^{-1}$). 
To represent absorption by the torus, we added an extra absorption 
component acting only on this power law. To allow for consistency 
between data and model, the fit requires the extra absorption component 
to have a column density of $\simgt 2\times 10^{24}$~cm$^{-2}$. About 
half of the Seyfert~2 population has torus column densities as large as 
this (e.g., Risaliti, Maiolino \& Salvati 1999).   
Many Seyfert~2s with large torus column densities show strong iron~K$\alpha$
emission lines. While we do not see evidence for iron~K$\alpha$ lines 
in our data, our constraints are not tight. For a narrow iron~K$\alpha$
line at 6.4~keV, our 90\% confidence upper limit on the equivalent width
is 750~eV. 


\section{Discussion, Conclusions, and Future Observations}

We have presented new \rosat\ HRI and \asca\ data that constrain
starburst and Seyfert activity in NGC~1672. Notably, we do not 
find any evidence for hard X-ray emission associated with the
nuclear source X-1 despite the claims for Seyfert activity made
based upon optical emission-line, infrared emission-line, X-ray,
and radio data (see BHI96 for a review). We also find no evidence
for long-term soft X-ray variability of X-1. If NGC~1672 indeed has a 
luminous Seyfert~2 nucleus, it must have been heavily obscured 
during the \asca\ observation by a `Compton-thick' torus. Our
X-ray spectral results stand in contrast to those of 
Awaki \& Koyama (1993), who claimed to detect luminous 
power-law emission from an obscured Seyfert nucleus with
$N_{\rm H}<3.2\times 10^{22}$~cm$^{-2}$. While it is perhaps 
possible that the torus column density along our line of sight 
has increased by a factor of $\simgt 60$ between the \ginga\ and 
\asca\ observations, the simplest explanation for the discrepant
results is that the \ginga\ data suffered from source confusion. 
The other arguments for Seyfert activity in NGC~1672, while entirely 
plausible, lack robustness. The reasoning based upon the composite
nature of the nuclear optical emission lines (V\'eron et~al. 1981) 
relies upon fairly subtle effects, and we note that 
Storchi-Bergmann et~al. (1996) have recently found that the nuclear
emission-line ratios can be explained via ionization by hot
($\approx 45\,000$~K) stars. In this case NGC~1672 may be an example 
of a non-AGN LINER. Furthermore, Moorwood \& Oliva (1988)
failed to detect the broad Brackett $\gamma$ line claimed by 
Kawara, Nishida \& Gregory (1987), and we are not aware of any
confirmation of the radio variability suggested by Tovmassian~(1968). 
While we must still admit the possibility of a luminous, highly
obscured Seyfert nucleus in NGC~1672, the physical interpretation
of its optical emission lines is not yet clear, and the balance
of evidence appears to be shifting away from this. 

The most interesting extranuclear source in NGC~1672 is X-3. We have
demonstrated that X-3 shows large-amplitude soft X-ray variability
on a timescale of $\approx 5$~years, suggesting that it is a single
(or at most a few) object. X-3 has a peak observed soft X-ray 
luminosity of $\approx 2.5\times 10^{39}$~erg~s$^{-1}$, and it probably
has an even larger hard X-ray luminosity. If the hard X-ray emission
of \S3.2 and Figure~4b originates from X-3 (the most probable 
possibility in our opinion), its 0.2--8~keV X-ray luminosity 
is $\simgt 6\times 10^{39}$~erg~s$^{-1}$. The fairly large ratio of 
hard to soft X-ray flux could be explained if X-3 suffers from some 
absorption within NGC~1672; the hard spectral components of
starburst galaxies are often absorbed by column densities of
$\approx 10^{22}$~cm$^{-2}$ (e.g., Ptak et~al. 1999).   
The variability, luminosity, and extranuclear location of X-3 suggest 
that it is a powerful X-ray binary, young supernova remnant, or
young hypernova remnant. No supernovae 
are listed for NGC~1672 in the Asiago Supernova Catalogue
(Barbon et~al. 1999). If X-3 is an accreting compact object that emits 
isotropically, the mass required by the the Eddington limit is 
$\simgt 45$~M$_\odot$. To our knowledge, the only source in our
Galaxy with a comparable X-ray luminosity is GRS~1915+105 
(e.g., Greiner, Morgan \& Remillard 1998), and X-3 ranks among the
more luminous of the `super-Eddington sources' seen in nearby galaxies
(e.g., Fabbiano 1998; Colbert \& Mushotzky 1999; and references therein). 
The new X-ray sources we find in NGC~1672 (X-7, X-8 and X-9) are 
also quite luminous and exceed the Eddington limit for 
a 1.4~M$_\odot$ object. They too are probably X-ray binaries or 
supernova remnants, and sources with comparable luminosities 
are seen fairly frequently in nearby 
spiral galaxies (e.g., Read, Ponman \& Strickland 1997, 
hereafter RPS97). 
We have compared the number and the luminosity distribution 
of our extranuclear X-ray sources with those 
of the nearby galaxies studied by RPS97, and NGC~1672 shows a 
moderately large although not extreme amount of extranuclear X-ray 
source activity. There are almost certainly many more X-ray binaries 
and supernova remnants lying below our current X-ray sensitivity 
threshold ($\approx 3\times 10^{38}$~erg~s$^{-1}$). 

A moderate-length \chandra\ observation holds particular promise for 
improving our understanding of NGC~1672. Separation 
of the diffuse nuclear X-ray emission 
from any point-like central source would provide a sensitive probe for 
low-level Seyfert activity. In addition, \chandra\ would allow an 
excellent imaging study of the nuclear starburst X-rays, and like the 
radio emission the X-rays might trace the nuclear ring  
(e.g., Lindblad \& J\"ors\"ater, in preparation). Furthermore, \chandra\ 
would provide much better spectral constraints upon X-3 and would 
detect X-ray binaries and supernova remnants down to 
$\approx 5\times 10^{37}$~erg~s$^{-1}$ in this nearby starburst galaxy. 
 

\acknowledgments

We thank M. Corcoran, H. Ebeling, E. Gotthelf and an anonymous referee
for helpful discussions, and we thank H. Ebeling for the use of his 
{\sc idl} software. We gratefully acknowledge financial support from 
NASA grants NAG5-4826 and NAG5-6023 (PJdN), and
NASA LTSA grant NAG5-8107 (WNB). 



\clearpage


\begin{deluxetable}{ccccccccc}
\tablenum{1}
\tablewidth{0pt}
\tablecaption {\rosat\ HRI Sources Coincident with NGC~1672}
\scriptsize
\tablehead{ 
\colhead{Source}                &
\colhead{}                      &
\colhead{}                      &
\colhead{}                      &
\colhead{Source}                &
\colhead{}                      &
\colhead{Mean Count}            &
\colhead{$F_{\rm X}/(10^{-14}$} &
\colhead{$L_{\rm X}/(10^{39}$}\\          
\colhead{Name}                      &
\colhead{Image}                     &
\colhead{$\alpha_{2000}$}           &
\colhead{$\delta_{2000}$}           &
\colhead{Counts}                    &
\colhead{$\sigma$}                  &
\colhead{Rate/($10^{-4}$ s$^{-1}$)} &
\colhead{erg cm$^{-2}$ s$^{-1}$)}   &
\colhead{erg s$^{-1}$)}             
}
\startdata
X-1             &
1992            &
04 45 42.2      &
$-59$ 14 50.3   & 
$174\pm 14.3$   &
22.4            &
$71.0\pm 5.83$  &
20.9            &
7.4             \nl
                &
1997            &
04 45 41.9      &
$-59$ 14 50.0   &
$234\pm 16.9$   &
24.5            &
$66.7\pm 4.8$   &
19.6            &
7.0             \nl
                &
Comb.           &
04 45 42.1      &
$-59$ 14 50.3   &
$400\pm 21.9$   &
33.1            &
$67.2\pm 3.7$   &
19.8            &
7.0             \nl
X-2             &
1992            &
04 45 53.0      &
$-59$ 14 56.5   &
$40\pm 7.4$     &
10.1            &
$16\pm 3.0$     &
6.6             &
2.6             \nl
                &
1997            &
04 45 52.5      &
$-59$ 14 55.0   &
$45\pm 8.3$     &
11.6            &
$13\pm 2.4$     &
5.2             &
2.1             \nl
                &
Comb.           &
04 45 52.8      &
$-59$ 14 55.4   &
$83\pm 11$      &
14.6            &
$14\pm 1.8$     &
5.6             &
2.2             \nl
X-3             &
1992            &
04 45 33.9      &
$-59$ 14 41.3   &
$22\pm 5.6$     &
7.7             &
$8.9\pm 2.3$    &
3.6             &
1.4             \nl
                &
1997            &
04 45 33.6      &
$-59$ 14 42.0   &
$54\pm 8.3$     &
9.9             &
$15\pm 2.4$     &
6.2             &
2.5             \nl
                &
Comb.           &
04 45 33.7      &
$-59$ 14 42.1   &
$78\pm 10$      &
12.8            &
$13\pm 1.7$     &
5.3             &
2.1             \nl
X-5             &
1992            &
04 45 49.6      &
$-59$ 12 49.0   &
$16\pm 5.0$     &
5.8             &
$6.7\pm 2.0$    &
2.7             &
---             \nl
                &
1997            &
---             &
---             &
$<13$           &
$<3.0$          &
$<3.8$          &
$<1.6$          &
---             \nl
                &
Comb.           &
04 45 49.6      &
$-59$ 12 49.2   &
$21\pm 6.7$     &
6.1             &
$3.5\pm 1.1$    &
1.4             &
---             \nl
X-7             &
1992            &
04 45 31.8      &
$-59$ 14 54.8   &
$12\pm 4.4$     &
4.7             &
$4.7\pm 1.8$    &
2.0             &
0.77            \nl
                &
1997            &
04 45 31.0      &
$-59$ 14 54.0   &
$20\pm 5.8$     &
6.6             & 
$5.8\pm 1.6$    & 
2.3             &
0.94            \nl
                &
Comb.           &
04 45 31.3      &
$-59$ 14 53.7   &
$30\pm 7.1$     &
7.4             &
$5.0\pm 1.2$    &
2.1             &
0.82            \nl
%
%
X-8             &
1992            &
---             &
---             &
$<9$            &
$<3.0$          &
$<3.9$          & 
$<1.6$          & 
$<0.64$          \nl
                &
1997            &
04 45 57.1      &
$-59$ 14 54.0   &
$7\pm 4.3$      &
4.0             &
$2.1\pm 1.2$    &
0.86            &
0.35            \nl
                &
Comb.           &
04 45 57.3      &
$-59$ 14 54.9   &
$<10$           &
$<3.0$          &
$<2.6$          &
$<1.1$         &
$<0.43$         \nl
X-9             &
1992            &
---             &
---             &
$<9$            &
$<3.0$          &
$<3.7$          & 
$<1.4$          & 
$<0.60$         \nl
                &
1997            &
---             &
---             &
$<12$           &
$<3.0$          &
$<3.3$          &
$<1.3$          &
$<0.55$         \nl
                &
Comb.           &
04 45 29.5      &
$-59$ 13 27.3   &
$10\pm 5.2$     &
4.0             &
$1.7\pm .9$     &
0.70            &
0.29            \nl
\enddata
\tablenotetext{}{The 1992 observation had an exposure of 24.5~ks, and
the 1997 observation had an exposure of 35.0~ks; the exposure time for
the combined (`Comb.') observation is 59.5~ks. 
Source positions have errors of $\approx 5^{\prime\prime}$. 
Source counts, source significances ($\sigma$), and source mean count rates 
are calculated using HRI channels 3--8 (see \S2.1; we have corrected for 
this channel range when calculating fluxes and luminosities).
$F_{\rm X}$ and $L_{\rm X}$ are given for the 0.2--2.0~keV band. $L_{\rm X}$ 
is corrected for Galactic absorption but $F_{\rm X}$ is not.
Our source fluxes and luminosities differ from those of BHI96 because we
use a narrower energy band (0.2--2.0~keV versus 0.1--2.5~keV) and a 
smaller distance (16.3~Mpc versus 22.8~Mpc). In addition, note that we 
have resolved source X-3 of BHI96 into two sources, X-3 and X-7. 
We have not computed $L_{\rm X}$ for X-5 because we believe this source is
associated with a foreground star (see \S2.2).
X-8 is detected at $3.8\sigma$ in the combined observation.}
\end{deluxetable}


\clearpage


\begin{figure}
\epsscale{0.5}
\plotfiddle{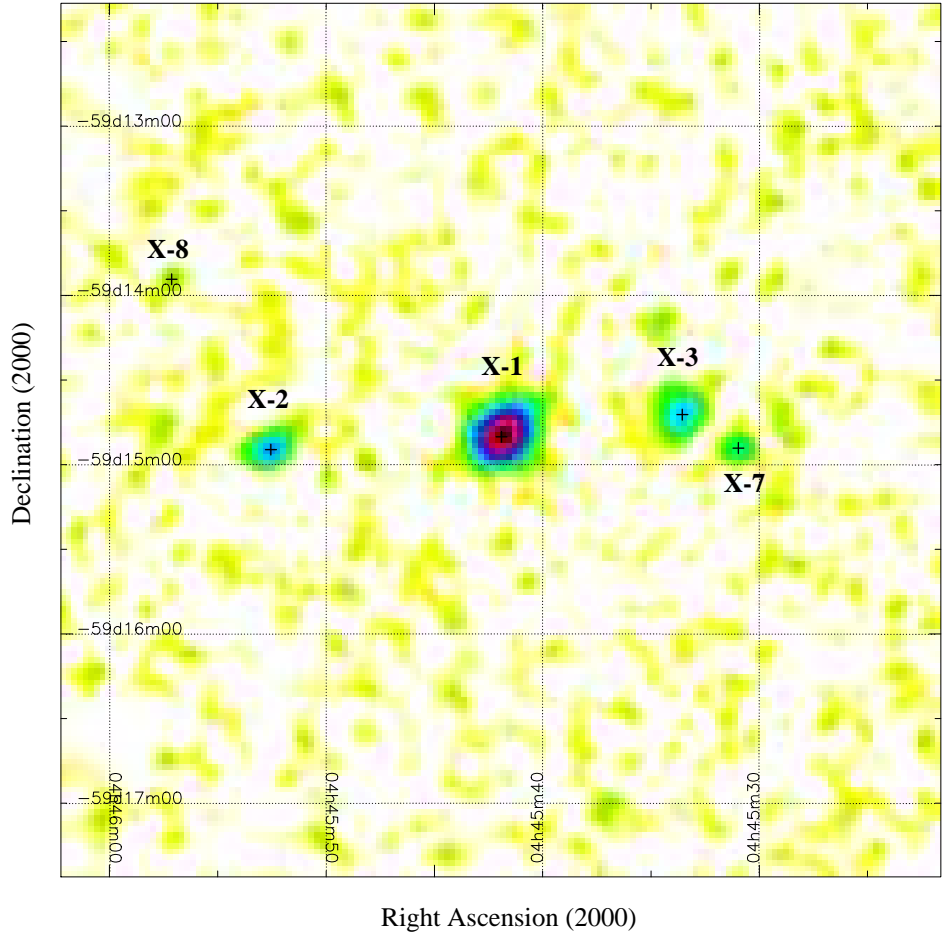}{340pt}{0}{70}{70}{-210}{-120}
\vspace{1.0in}
\caption{\pss\ significance map (see \S2.2) for the 1997 HRI observation of 
NGC~1672. The crosses on the image mark sources that are detected at the
$\geq 4\sigma$ level. X-3 and X-7 are clearly separated, and the 
probable new source X-8 is identified.
\label{fig1}}
\end{figure}



\begin{figure}
\epsscale{0.5}
\plotfiddle{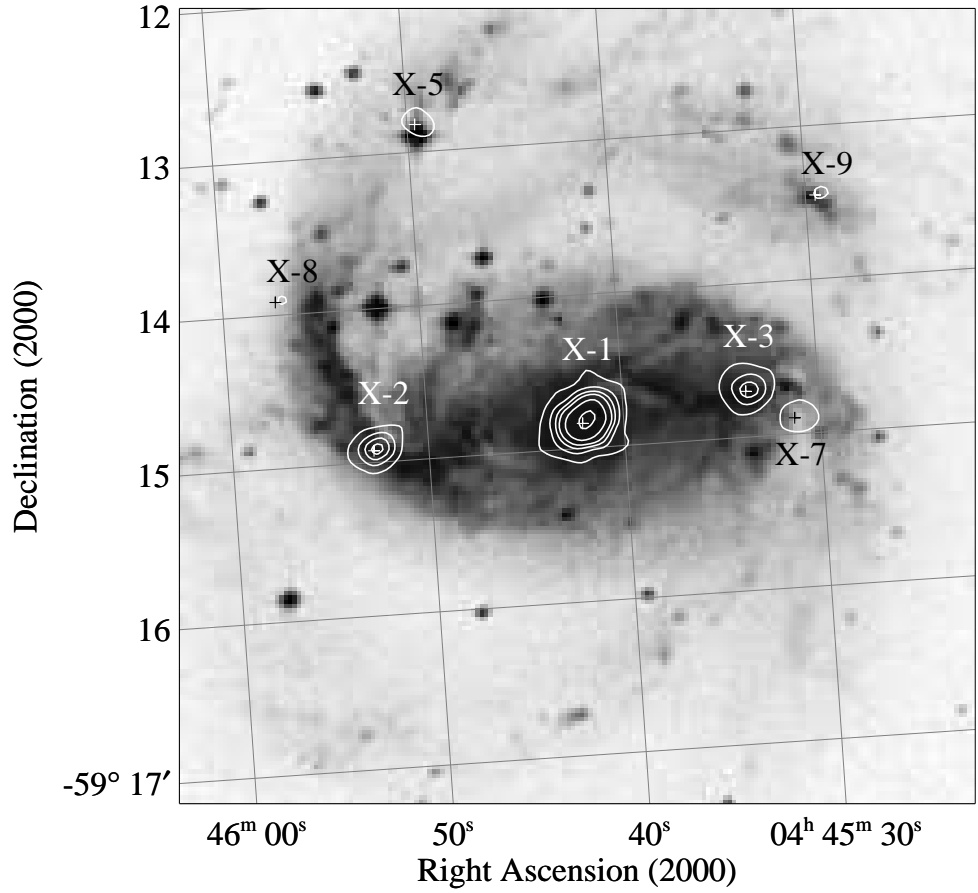}{340pt}{0}{70}{70}{-210}{-120}
\vspace{0.5in}
\caption{Contours of an iteratively smoothed `combined' HRI image 
(see \S2.1) overlaid on the image of NGC~1672 from the UK Schmidt
southern sky survey $J$ plate. We have labeled the $\geq4\sigma$ sources,
and contours are at 8.8, 17.6, 26.5, 35.4, 53.0 and 88.4\% of the maximum 
pixel value (see Table~1 for absolute source fluxes). 
\label{fig2}}
\end{figure}



\begin{figure}
\epsscale{0.5}
\vspace{-1.2in}
\plotfiddle{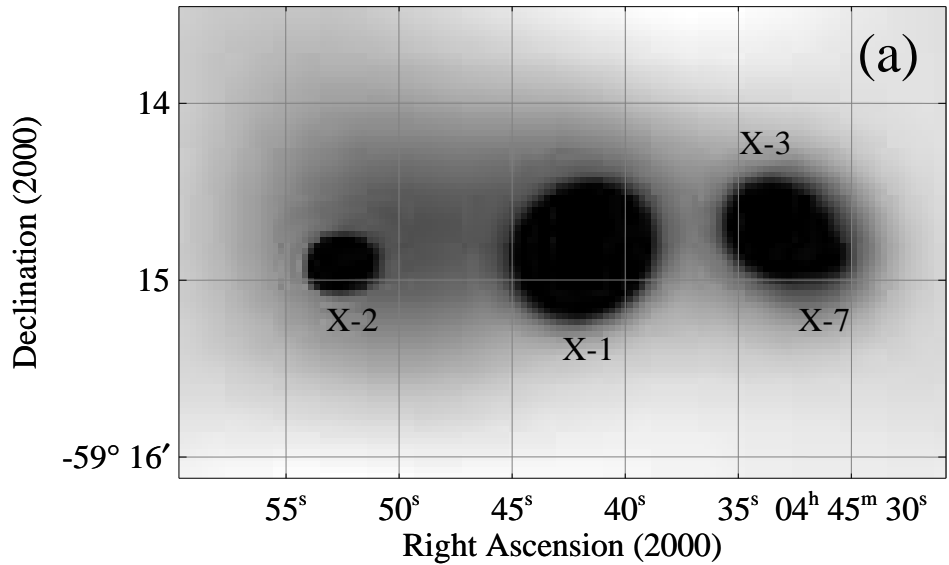}{340pt}{0}{70}{70}{-210}{-120}
\plotfiddle{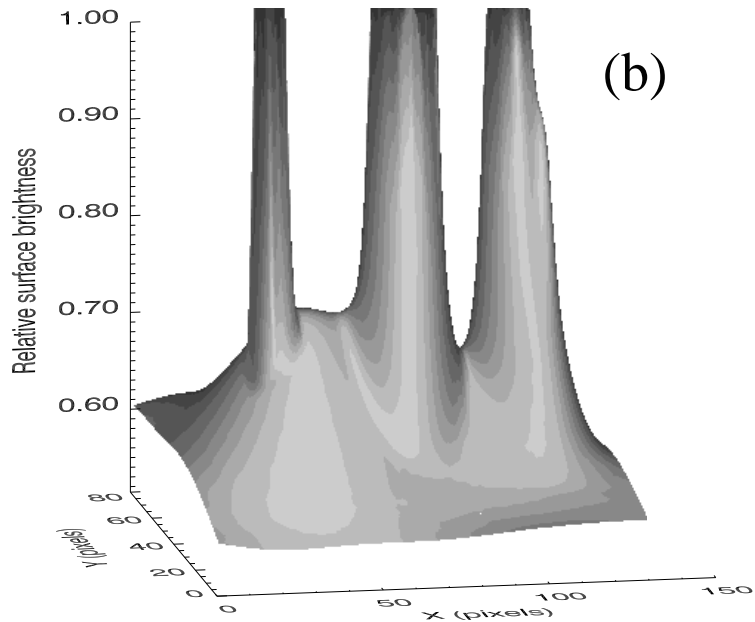}{340pt}{0}{70}{70}{-210}{-120}
\vspace{-1.9in}
\caption{(a) Adaptively smoothed `combined' HRI image (see \S2.1)
showing the diffuse soft X-ray emission from the region between 
X-1 and X-2. 
(b) Three-dimensional representation of the adaptively smoothed 
`combined' HRI image. From left to right, the three strong sources 
are X-2, X-1 and X-3. Note the diffuse emission between X-1 and X-2. 
The surface brightness in the region between X-1 and X-3 is lower 
than that between X-1 and X-2 (even though X-1 and X-3 are closer 
together). 
\label{fig3}}
\end{figure}




\begin{figure}
\epsscale{0.5}
\vspace{-1.2in}
\plotfiddle{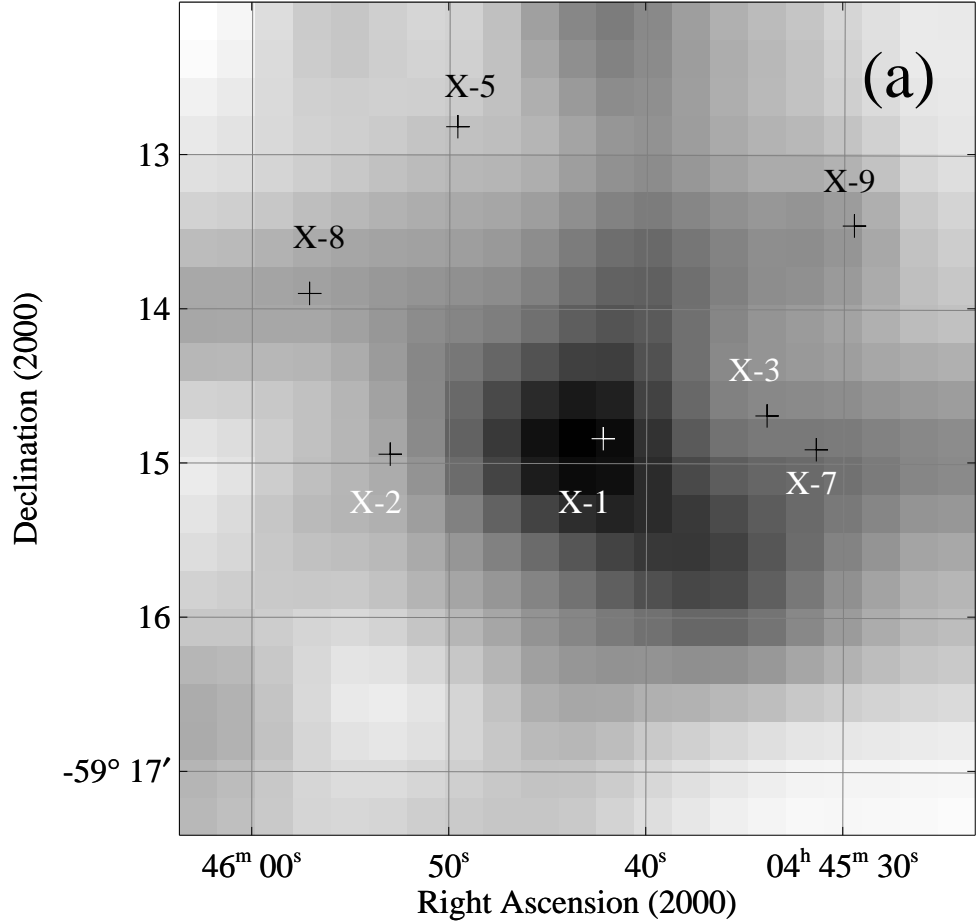}{340pt}{0}{70}{70}{-210}{-120}
\vspace{0.4in}
\caption{(a) Iteratively smoothed \asca\ GIS2+GIS3 image in the `soft' 1.0--3.0~keV band. 
The sources detected by the \rosat\ HRI are marked and labeled for comparison.
Note that the dominant emission seen by \asca\ is coincident with X-1. 
(b) Iteratively smoothed \asca\ GIS2+GIS3 image in the `ultra-hard' 5--10~keV band. 
Note that the dominant emission seen by \asca\ is coincident with X-3 and X-7. 
\label{fig4a}}
\end{figure}



\begin{figure}
\epsscale{0.5}
\plotfiddle{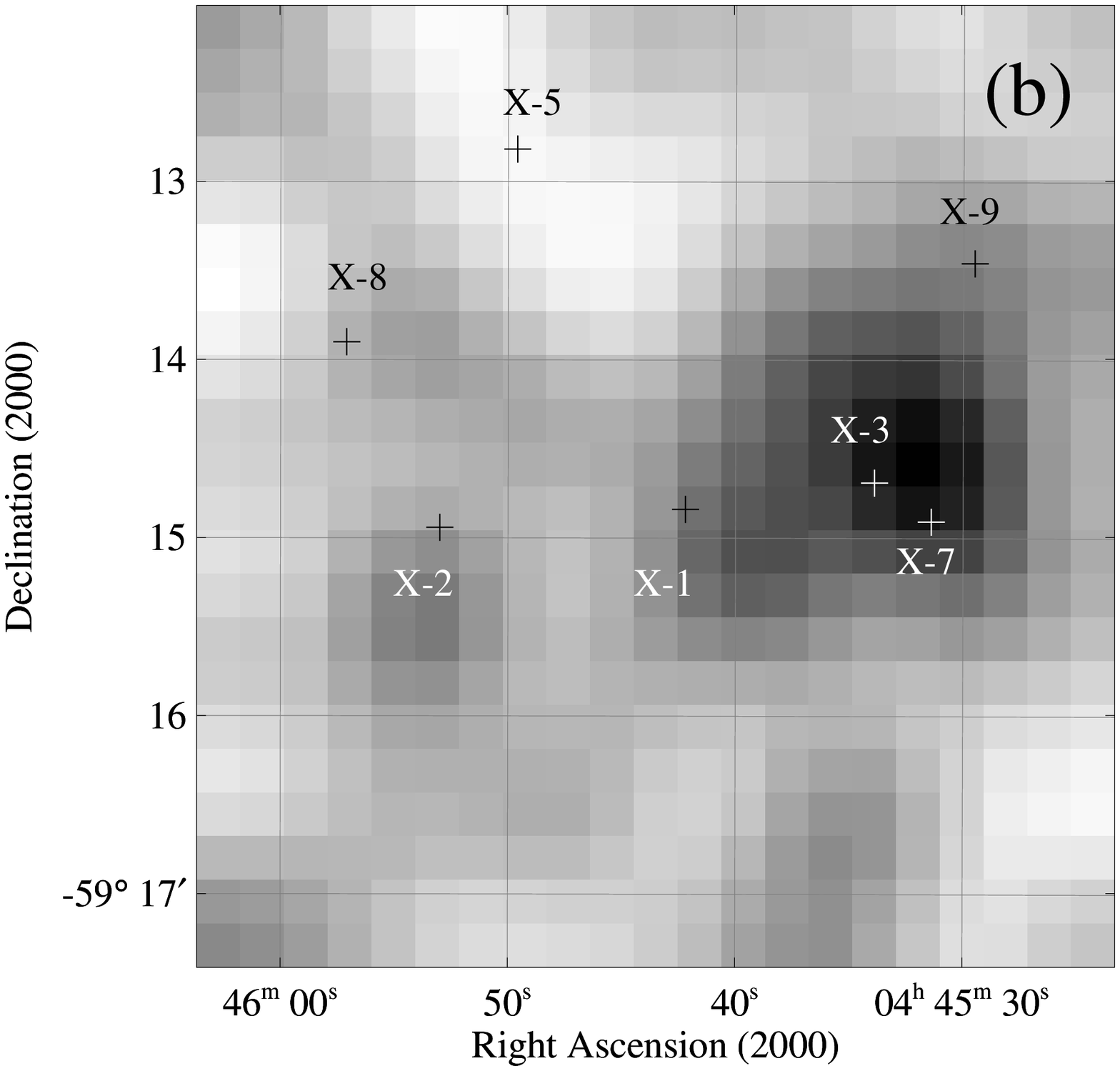}{340pt}{0}{70}{70}{-210}{-120}
\end{figure}



\begin{figure}
\epsscale{0.5}
\plotfiddle{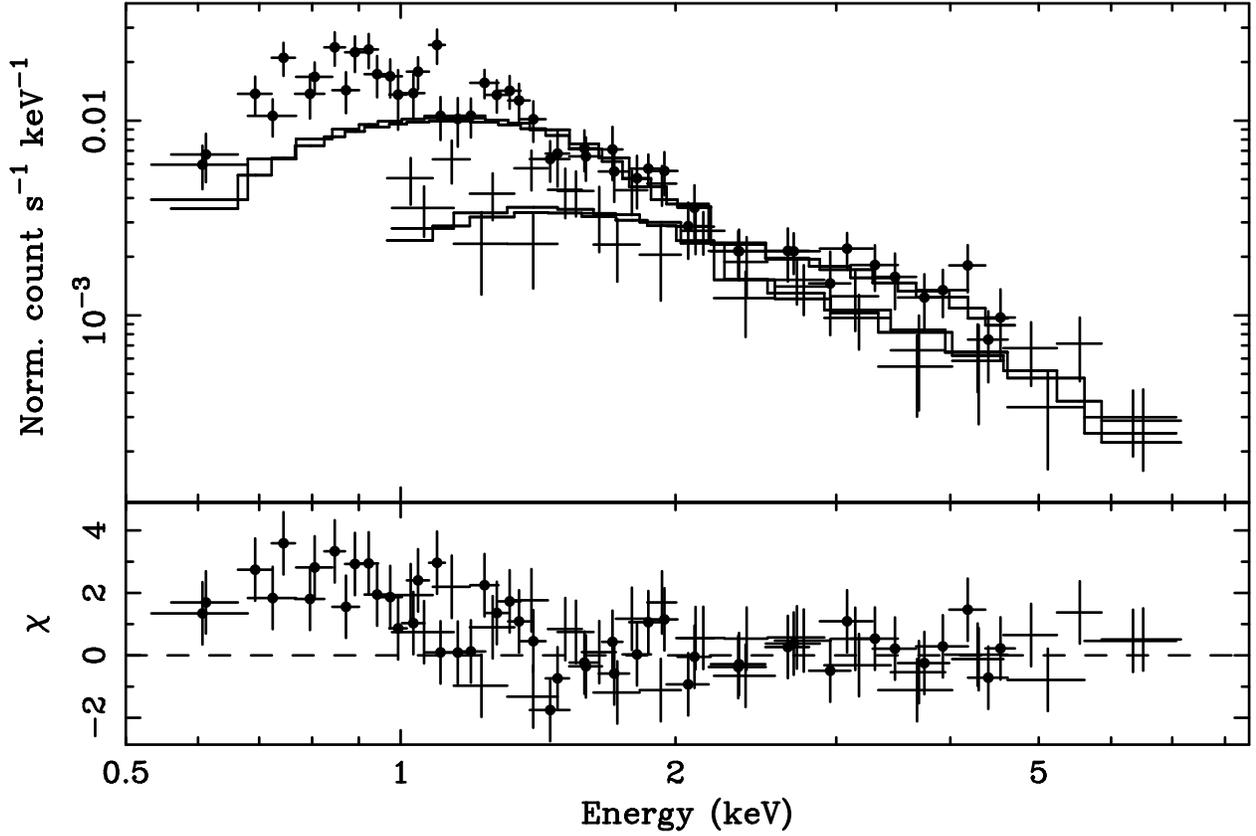}{340pt}{-90}{70}{70}{-260}{430}
\vspace{0.4in}
\caption{\asca\ SIS (solid dots) and GIS (plain crosses) spectra of NGC~1672. 
A power-law model has been fit to the data above 2~keV and then
extrapolated downward in energy to show the deviations from a power law.
The ordinate for the lower panel (labeled $\chi$) shows the fit residuals
in terms of sigmas with error bars of size unity. Note the systematic 
positive residuals at low energies. 
\label{fig5}}
\end{figure}




\begin{thebibliography}{}

\bibitem[]{}
Allan, D. J. 
1995, 
{\sc asterix} User Note 004: Source Searching and Parameterisation.
University of Birmingham, Birmingham

\bibitem[]{}
Allan, D. J. \& Vallance, R. J.
1995, 
{\sc asterix} X-ray Data Processing System.
University of Birmingham, Birmingham

\bibitem[]{} 
Arnaud, K. A. 1996, in 
Astronomical Data Analysis Software and Systems V: ASP Conference Series 101, 
ed. Jacoby, G. \& Barnes, J. 
(ASP Press, San Francisco), p. 17  

\bibitem[]{} 
Awaki, H. \& Koyama, K.
1993,
Adv. Space Res., 13, 221

\bibitem[]{} 
Barbon, R., Buond\'\i, V., Cappellaro, E. \& Turatto, M.
1999, 
A\&AS, in press
(astro-ph/9908046)  

\bibitem[]{} 
Baumgart, C. W. \& Peterson, C. J.
1986, 
PASP, 98, 56

\bibitem[]{} 
Brandt, W. N., Halpern, J. P. \& Iwasawa, K. 
1996,
MNRAS, 281, 687 (BHI96)   

\bibitem[]{} 
Cash, W. 
1979, 
ApJ, 228, 939

\bibitem[]{} 
Colbert, E.J.M. \& Mushotzky, R.F.
1999, 
ApJ, 519, 89

\bibitem[]{}
David, L. P., Harnden, F. R., Kearns, K. E. \& Zombeck, M. V.
1999,
The \rosat\ High Resolution Imager Calibration Report.
U.S. \rosat\ Science Data Center, Cambridge

\bibitem[]{} 
Ebeling, H., White, D. A. \& Rangarajan, F. V. N.
1999,
MNRAS, submitted

\bibitem[]{} 
Evans, I. N., Koratkar, A. P, Storchi-Bergmann, T., Kirkpatrick, H., 
Heckman, T. M. \& Wilson, A. S.
1996, 
ApJS, 105, 93

\bibitem[]{}
Fabbiano, G.
1998, 
The Hot Universe: Proceedings of IAU Symposium 188,
ed. Koyama, K., Kitamoto, S. \& Itoh, M. 
(Kluwer, Dordrecht), p. 93

\bibitem[]{}
Giommi, P., Angelini, L. \& White, N.
1997, 
The {\sc ximage} Users' Guide: Version 2.53. 
NASA/GSFC, Greenbelt

\bibitem[]{}
Gotthelf, E. V.
1996,
\asca\ News, 4, 31 


\bibitem[]{}
Greiner, J., Morgan, E. H. \& Remillard, R. A.
1998, 
New Astronomy Reviews, 42, 597 (astro-ph/9806323)

\bibitem[]{}
Heiles, C. \& Cleary, M. N.
1979,
Aust. J. Phys. Astrophys. Suppl., 47, 1

\bibitem[]{}
Ingham, J. \& Arnaud, K. 
1998,
The {\sc xselect} Users' Guide. 
NASA/GSFC, Greenbelt

\bibitem[]{}
Iwasawa, K., Fabian, A. C. \& Nandra, K.
1999, 
MNRAS, in press (astro-ph/9904071)

\bibitem[]{}
Kawara, K., Nishida, M. \& Gregory, B.
1987, 
ApJ, 321, L35

\bibitem[]{}
Maccacaro, T., Gioia, I. M., Wolter, A., Zamorani, G. \& Stocke, J. T. 
1988, 
ApJ, 326, 680

\bibitem[]{}
Moorwood, A. F. M. \& Oliva, E.
1988, 
A\&A, 203, 278

\bibitem[]{}
Morse, J. A.
1994, 
PASP, 106, 675

\bibitem[]{}
Mukai, K. 
1997, 
The {\sc pimms} Users' Guide. 
NASA/GSFC, Greenbelt

\bibitem[]{}
Osmer, P. S., Smith, M. G. \& Weedman, D. W.
1974, 
ApJ, 192, 279

\bibitem[]{}
Pastoriza, M. G.
1967, 
The Observatory, 
87, 225

\bibitem[]{}
Pier, E. A. 
1997, 
\asca\ Getting Started Guide for Revision 2 Data: Version 6.1. 
NASA/GSFC, Greenbelt

\bibitem[]{}
Ptak, A., Serlemitsos, P., Yaqoob, T. \& Mushotzky, R.
1999, 
ApJS, 120, 179

\bibitem[]{}
Read, A. M., Ponman, T.J. \& Strickland, D. K. 
1997, 
MNRAS, 286, 626 (RPS97)

\bibitem[]{}
Risaliti, G., Maiolino, R. \& Salvati, M.
1999, 
ApJ, in press (astro-ph/9902377)

\bibitem[]{}
Storchi-Bergmann, T., Wilson, A. S. \& Baldwin, J. A.
1996,
ApJ, 460, 252 

\bibitem[]{}
Tanaka, Y., Inoue, H. \& Holt, S. S. 
1994, 
PASJ, 46, L37

\bibitem[]{}
Tovmassian, H.M. 
1968, 
The Observatory, 88, 227

\bibitem[]{} 
V\'eron, M. P., V\'eron, P. \& Zuiderwijk, E. J.
1981,
A\&A, 98, 34

\end{thebibliography}
\end{document}